\documentclass[conference,twocolumn]{IEEEtran}

\usepackage{amsmath, amssymb,bm}
\usepackage{bbm}
\usepackage{url}
\usepackage{graphicx}
\usepackage{cite}
\usepackage{esvect}
\usepackage{subcaption}
\usepackage{amsthm}

\usepackage[ruled,linesnumbered]{algorithm2e} 
\usepackage{caption}

\allowdisplaybreaks[1]
\allowdisplaybreaks
\usepackage{comment}
\usepackage{pifont}
\usepackage{color}

\ifCLASSINFOpdf
\else
\fi

\begin{document}
\title{Towards Reconfigurable Intelligent Surfaces Powered Green Wireless Networks
}
%
%
%

\author{\IEEEauthorblockN{Siyuan Sun\IEEEauthorrefmark{1}\IEEEauthorrefmark{2}\IEEEauthorrefmark{3}, Min Fu\IEEEauthorrefmark{1}, Yuanming Shi\IEEEauthorrefmark{1}, and Yong Zhou\IEEEauthorrefmark{1}}
	\IEEEauthorblockA{\IEEEauthorrefmark{1}School of Information Science and Technology, ShanghaiTech University, Shanghai 201210, China
	}
	\IEEEauthorblockA{\IEEEauthorrefmark{2}Shanghai Institute of Microsystem and Information Technology, Chinese Academy of Sciences, China
	}
	\IEEEauthorblockA{\IEEEauthorrefmark{3}University of Chinese Academy of Sciences, Beijing 100049, China
	}
	E-mail:  \{sunsy, fumin, shiym, zhouyong\}@shanghaitech.edu.cn
}

\maketitle

\thispagestyle{empty}

\IEEEpeerreviewmaketitle

\begin{abstract}
The adoption of reconfigurable intelligent surface (RIS) in wireless networks can enhance the spectrum- and energy-efficiency by controlling the propagation environment. 
Although the RIS does not consume any transmit power, the circuit power of the RIS cannot be ignored, especially when the number of reflecting elements is large. 
In this paper, we propose the joint design of beamforming vectors at the base station, active RIS set, and phase-shift matrices at the active RISs to minimize the network power consumption, including the RIS circuit power consumption, while taking into account each user's target data rate requirement and each reflecting element's constant modulus constraint. 
However, the formulated problem is a mixed-integer quadratic programming (MIQP) problem, which is NP-hard. 
To this end, we present an alternating optimization method, which alternately solves second order cone programming (SOCP) and MIQP problems to update the optimization variables. 
Specifically, the MIQP problem is further transformed into a semidefinite programming problem by applying binary relaxation and semidefinite relaxation. Finally, an efficient algorithm is developed to solve the problem. 
Simulation results show that the proposed algorithm significantly reduces the network power consumption and reveal the importance of taking into account the RIS circuit power consumption. 
\end{abstract}


\section{Introduction}
With the upsurge of various mobile services, especially the services supported by artificial intelligence, the global mobile data traffic is expected to grow rapidly and reach $77$ exabytes per month by 2022 \cite{8808168}. 
To support those data-intensive mobile services, various technologies have been proposed to improve the capacity of the emerging fifth-generation (5G) wireless systems. 
In particular, millimeter wave (mmWave) communication, ultra-dense network (UDN), and massive multiple-input multiple-output (MIMO) are three most promising techniques that can significantly enhance the spectrum-efficiency \cite{Key2017}. 
However, the ultra-dense placement of base stations (BSs) with the massive antenna arrays, especially in mmWave networks, generally incurs excessive energy consumption \cite{wu2019towards}. 
According to China Mobile, the power consumption of one 5G BS is three times higher than that of one BS in long-term evolution (LTE). 
Thus, it is urgent to develop energy-efficient techniques to realize green communications. 

Deploying reconfigurable intelligent surfaces (RISs) in wireless networks has recently been recognized as a cost-effective way to enhance the spectrum-efficiency and energy-efficiency \cite{di2019smart, bjornson2019intelligent, basar2019large}. 
In particular, RIS is a flat man-made metasurface composed of multiple passive reflecting elements, each of which is capable of altering the phase shift of the incident signals in a programmable manner, thereby enhancing the received signal power at the receiver \cite{Liang2019Large, 8466374}. 
Without containing active radio frequency (RF) chains for power amplification, the RIS does not consume any transmit power. 
Furthermore, due to its small hardware footprint, the RIS can be flexibly deployed in both indoors (e.g., ceilings) and outdoors (e.g., buildings), as well as be integrated into the existing cellular systems \cite{8847342}.

The research on the beamforming design for RIS-assisted wireless networks has attracted considerable attention recently \cite{wu2018intelligent, fu2019intelligent, huang2019reconfigurable, Hua2019Reconfigurable, tang2019joint}. 
Specifically, the authors in \cite{wu2018intelligent} proposed the joint design of the active beamforming at the BS and the passive beamforming at the RIS to minimize the total transmit power of multiple-input single-output (MISO) wireless networks. 
The advantages of RIS were leveraged to reduce the transmit power of downlink non-orthogonal multiple access (NOMA) transmission in \cite{fu2019intelligent}. 
By taking into account the RIS circuit power, the authors in \cite{huang2019reconfigurable} formulated an energy-efficiency maximization problem, followed by proposing two efficient algorithms based on alternating optimization, gradient descent search, and fractional programming. 
Moreover, RIS was also utilized to reduce the energy consumption of edge inference systems in \cite{Hua2019Reconfigurable} and increase the minimum energy harvested among multiple devices 
in \cite{tang2019joint}. 
The aforementioned studies demonstrated that the BS transmit power can be reduced by deploying a single RIS. 


As the users are in general randomly distributed across the network coverage area, deploying multiple RISs has the potential to further enhance the network performance by increasing the probability of the signals being reflected. 
The authors in \cite{wu2019towards} discussed the potential of deploying large-scale RISs to alleviate the co-channel interference and suggested that the RIS should be deployed to ensure high-rank MIMO channels for achieving the spatial multiplexing gains. 
The authors in \cite{hu2018beyond} utilized multiple RISs for device positioning. 
The authors in \cite{di2019reflection} modeled the environmental objects coated with meta-surface with a random process and analytically derived the probability that one randomly distributed object can act as a reflector. 
In addition, multiple RISs were leveraged to enhance the efficiency of simultaneous wireless information and power transfer (SWIPT) in \cite{wu2019joint}. 
Although the BS transmit power can be reduced with the assistance of RISs, the deployment of multiple RISs incurs non-negligible RIS circuit power consumption, which is comparable to the transmit power of the BS.  
However, most of the existing studies except \cite{huang2019reconfigurable} did not taken into account the RIS circuit power consumption. 
To provide on-demand services in networks with dynamic traffic fluctuations and channel conditions, some RISs may be dynamically switched off to reduce the RIS circuit power consumption. 
However, only a single RIS that cannot be switched off to save energy was considered in \cite{huang2019reconfigurable}.


In this paper, we consider a \textit{multi-RIS-assisted} multi-user MISO system, where one BS serves multiple users with the assistance of multiple RISs. 
The main objective is to minimize the network power consumption, including the circuit power consumption of the RISs and the transmit power consumption of the BS, subject to the quality-of-service (QoS) constraints of the users and the constant modulus constraints of the RISs. 
The RIS circuit power consumption is determined by the set of active RISs, while the transmit power consumption of the BS can be minimized through optimizing the beamforming vectors of the BS and the phase-shift matrices of the active RISs. 
There exists a tradeoff between the RIS circuit power consumption and the BS transmit power consumption. 
To this end, we propose the joint design of the beamforming vectors at the BS, the active RIS set, and the phase-shift matrices at the active RISs. 

The formulated problem turns out to be a \textit{mixed-integer quadratic programming (MIQP)} problem, which is NP-hard. 
We present an alternating optimization framework that optimizes the BS beamforming as well as active RIS set and the corresponding phase shifts alternately. 
In each alternation, the beamforming optimization problem is recast as a second order cone programming (SOCP) problem, while the active RIS set and the corresponding phase-shift matrices optimization problem turns out to be a MIQP problem. 
In particular, we transform the MIQP problem into a non-convex quadratically constrained quadratic programming (QCQP) problem via binary relaxation, followed by recasting the resulting non-convex QCQP problem as a rank-constrained semidefinite programming (SDP) problem via matrix lifting. 
By utilizing the powerful semidefinite relaxation (SDR) technique, we propose an efficient algorithm to solve the problem. 
Simulation results show that the proposed algorithm significantly reduces the network power consumption and reveal the importance of taking into account the RIS circuit power consumption in designing RIS powered green wireless networks.

\begin{figure}[t] \centering
	\includegraphics[scale = 0.45]{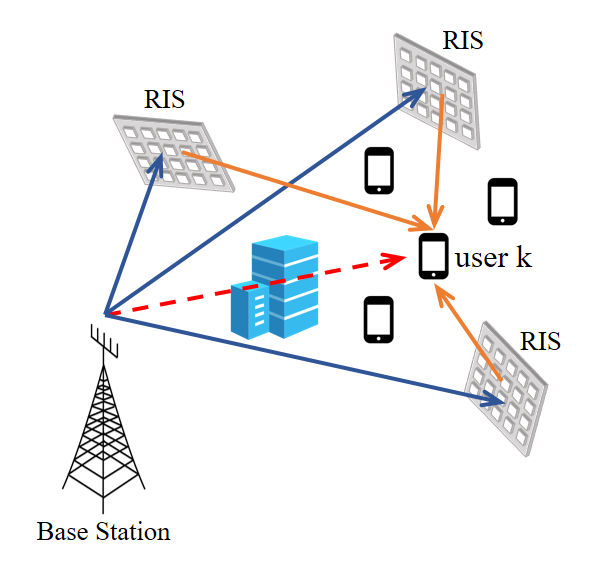}
	\caption{An illustration of a multi-RIS-assisted multi-user MISO downlink cellular network. }
	\label{Fig_system}
	\vspace{-3mm}
\end{figure}

\section{System Model and Problem Formulation} \label{Sec_SysMod}
\subsection{System Model}
Consider the downlink transmission of a multi-RIS-assisted cellular network, where one $M$-antenna BS serves $K$ single-antenna users with the assistance of $L$ distributed RISs, as shown in Fig. \ref{Fig_system}. 
We denote $\mathcal{K} = \{ 1, 2, \ldots, K\}$ and $\mathcal{L} = \{ 1, 2, \ldots, L \}$ as the sets of user and RIS indices, respectively. 
The number of passive reflecting elements on RIS $l$ is denoted as $N_l, \forall \, l \in \mathcal{L}$. 
Each user is expected to receive a unicast signal from the BS. 
With linear precoding, the signal transmitted by the BS can be expressed as $
\bm x  = \sum_{k\in\mathcal{K}} \bm \omega_k s_k$, 
where $s_k \in \mathbb{C}$ and $\bm \omega_k\in \mathbb{C}^{M \times 1}$ denote the complex valued information symbol and the beamforming vector for user $k$, respectively. Without loss of generality, we assume that signals $\{s_k|k \in \mathcal{K}\}$ are independent, and have zero mean and unit variance, i.e., $\mathbb{E} [s_k s_k^{\mathrm{H}} ] = 1$, and $\mathbb{E} [s_k s_j^{\mathrm{H}} ] =0, \forall \, k \neq j$, where $(\cdot)^{\mathrm{H}}$ denotes the conjugate transpose. 
Besides, the BS has a maximum transmit power constraint $\sum_{k\in \mathcal{K}} \| \bm \omega_k \|^2 \le P_{\mathrm{max}}$, 
where $P_{\mathrm{max}}$ denotes the maximum transmit power of BS. 

With the assistance of RIS, the transmit power of the BS can be significantly reduced \cite{wu2018intelligent}. 
However, the circuit power consumption of the RIS is non-negligible, especially when the number of reflecting elements is large, as demonstrated in \cite{huang2019reconfigurable}. 
Hence, it becomes critical to dynamically switch off some RISs to reduce the energy consumption while the QoS requirements of all users can be satisfied. 
We denote $\mathcal{A} \subseteq \mathcal{L}$ as the index set of active RISs and $\mathcal{Z} \subseteq \mathcal{L}$ as the index set of inactive RISs with $\mathcal{A} \cup \mathcal{Z} = \mathcal{L}$. 
In addition, we denote the channel responses from the BS to RIS $l$, from RIS $l$ to user $k$, and from the BS to user $k$ as $\bm G_l \in \mathbb{C}^{N_l \times M}$, $\bm h_{l,k} \in \mathbb{C}^{N_l \times 1}$, and $\bm g_{k} \in \mathbb{C}^{M \times 1}, \forall \, l \in \mathcal{L}, \forall \, k \in \mathcal{K}$, respectively. 
We consider quasi-static block fading, i.e., all the channel responses remain invariant during one transmission block. 
In addition, we assume that perfect channel state information (CSI) is available at the BS, as in \cite{wu2018intelligent, fu2019intelligent, huang2019reconfigurable, Hua2019Reconfigurable, tang2019joint}. 
The baseband signal, transmitted from the BS and reflected by the active RISs in set $\mathcal{A}$, received at user $k \in \mathcal{K}$ can be expressed as 
\begin{eqnarray}
y_k = \left( \sum_{l \in \mathcal{A}} \bm h_{l,k}^{\mathrm{H}} \bm \Theta_l \bm G_l + \bm g_k^{\mathrm{H}} \right) \sum_{i\in \mathcal{K}} \bm \omega_i s_i + z_k, \forall \, k, 
\end{eqnarray}
where $\bm \Theta_l = \mathrm{diag}( \rho_le^{j\varphi_{l,1}}, \rho_le^{j\varphi_{l,2}} \ldots, \rho_le^{j\varphi_{l,N_l}})\in \mathbb{C}^{N_l \times N_l}$ represents the diagonal phase-shift matrix of RIS $l$, $\rho_l$ denotes the amplitude reflection coefficient of RIS $l$, $\varphi_{l,n}\in[0,2\pi)$ denotes the phase shift of passive reflecting element $n$ equipped on RIS $l$, and $z_k \sim \mathcal{CN}(0, \sigma_k^2)$ denotes the additive white Gaussian noise (AWGN) with $\sigma_k^2$ being the noise power of user $k$. 
As in \cite{wu2018intelligent, fu2019intelligent, huang2019reconfigurable, Hua2019Reconfigurable, tang2019joint}, we assume that the power of the signals reflected by the RISs two or more times is negligible. 

By assuming that each user adopts the single-user detection strategy, the achievable signal-to-interference-plus-noise ratio (SINR) at user $k \in \mathcal{K}$ can be expressed as 
\begin{eqnarray} \label{Eq_SINRk3}
\mathrm{SINR}_{k} (\mathcal{A}) \!=\! \frac{\left|\left( \sum_{l \in \mathcal{A}} \bm h_{l,k}^{\mathrm{H}} \bm \Theta_l \bm G_l + \bm g_k^{\mathrm{H}} \right) \bm \omega_k \right|^2}{\sum_{j \neq k} \left|\left( \sum_{l \in \mathcal{A}} \bm h_{l,k}^{\mathrm{H}} \bm \Theta_l \bm G_l + \bm g_k^{\mathrm{H}} \right)\bm \omega_j \right|^2 \!\!\!+ \! \sigma_k^2}. 
\end{eqnarray}


\subsection{Power Consumption Model}
The power consumption model is critical for the design of a green wireless network. 
In particular, the total power consumption of an RIS-assisted wireless network includes the BS transmit power, the BS circuit power, and the RIS circuit power. 
It is worth noting that each RIS, without the active RF chains for power amplification, does not consume any transmit power because of the passive nature of the reflecting elements. 
By adopting an empirical linear power consumption model for the BS, the total power consumption is defined as 
\begin{eqnarray} \label{Eq_NPC}
\hat{p}(\mathcal{A}, \{ \bm \omega_k \}) = \sum_{k\in \mathcal{K}} \frac{1}{\eta} \| \bm \omega_k \|^2 + P_{\mathrm{BS}} + \sum_{l \in \mathcal{A}}P_{\mathrm{RIS}}(N_l),
\end{eqnarray}
where $\eta$ is the drain efficiency of the RF power amplifier at the BS, $P_{\mathrm{BS}}$ denotes the circuit power consumption of the BS, and $P_{\mathrm{RIS}}(N_l)$ denotes the circuit power consumption of the RIS $l$ with $N_l$ passive reflecting elements. 

According to \cite{Huang2018Energy}, the circuit power consumption of the RIS depends on the resolution of each reflecting element. 
In particular, the typical values of the power consumption of each reflecting element with $3$-bit, $4$-bit, $5$-bit, and $6$-bit phase-shift resolutions are $1.5$ mW, $4.5$ mW, $6$ mW, and $7.8$ mW, respectively. 
With a large number of reflecting elements on each RIS, the circuit power consumption of the RIS cannot be ignored. 
For tractability, we consider the RIS with continuous phase shifting in this paper. 
By denoting $P_{\mathrm{RE}}$ as the power consumption of each reflecting element, the power consumption of an RIS with $N_l$ passive elements is $P_{\mathrm{RIS}}(N_l) = N_l P_{\mathrm{RE}},\forall \, l \in \mathcal{A}$. 

As the circuit power consumption of the BS is a constant, minimizing the total network power consumption $\hat{p}(\mathcal{A}, \{ \bm \omega_k \})$ in (\ref{Eq_NPC}) is equivalent to minimizing the following re-defined network power consumption\begin{eqnarray} \label{Eq_NPC1}
p(\mathcal{A}, \{ \bm \omega_k \}) = \sum_{k \in \mathcal{K}} \frac{1}{\eta} \| \bm \omega_k \|^2 + \sum_{l \in \mathcal{A}}P_{\mathrm{RIS}}(N_l), 
\end{eqnarray}
where both the BS transmit power and RIS circuit power consumption are considered. 

\subsection{Problem Formulation}
The network power consumption model given in (\ref{Eq_NPC1}) indicates that the power consumption can be minimized by reducing the transmit power of the BS and decreasing the number of active RISs. 
However, there exists a tradeoff between these two strategies. 
On one hand, to reduce the transmit power of the BS, more RISs are required to be active to exploit the passive array gain. 
On the other hand, increasing the number of active RISs incurs a higher RIS circuit power consumption.  

To this end, we propose a joint design of the beamforming vectors at the BS, the active RIS set, and the phase-shift matrices at the active RISs to minimize the network power consumption in (\ref{Eq_NPC1}), while taking into account the target SINR requirement of each user and the constant modulus constraint of each passive reflecting element. 
To facilitate the problem formulation, we denote $\theta_{l,n} = \rho_l e^{j\varphi_{l,n}}, \forall \, l \in \mathcal{A}, \forall \, n \in \mathcal{N}_l = \{ 1, 2, \ldots, N_l\}$. As a result, the phase-shift matrix can be rewritten as $\bm \Theta_l = \mathrm{diag}( \theta_{l,1}, \theta_{l,2} \ldots, \theta_{l,N_l})\in \mathbb{C}^{N_l \times N_l}, \forall \, l \in \mathcal{A}$. 
The formulated network power minimization problem is given by
\begin{subequations} \label{OP_Original}
\begin{align}
\hspace{-3mm} \mathcal{P}: \mathop{\text{minimize}}_{ \mathcal{A}, \{ \bm \omega_k \}, \{ \bm \Theta_l \}} \label{P_Obj}\hspace{3mm}  & p(\mathcal{A}, \{\bm  \omega_k \})  \\
\text{subject to} 
\label{P_SINR} \hspace{3mm}  & \mathrm{SINR}_k(\mathcal{A}) \ge \gamma_k, \forall \, k \in \mathcal{K},  \\
\label{P_power} \hspace{3mm}  & \sum_{k\in \mathcal{K}} \| \bm \omega_k \|^2 \le P_{\mathrm{max}},  \\
\label{P_unit} \hspace{3mm}  & | \theta_{l,n}| = \rho_l, \forall \, l \in \mathcal{A},\forall \, n \in \mathcal{N}_l,
\end{align}
\end{subequations}
where $\gamma_k$ denotes the SINR threshold of signal $s_k, \forall \, k \in \mathcal{K}$. 

Problem $\mathcal{P}$ is a joint BS beamformer, active RIS set, and RIS phase shifts optimization problem.  
It turns out to be an MIQP problem, which is intractable to be solved in general and imposes the following three challenges. 
First, the objective function (\ref{P_Obj}) is combinatorial due to the RIS set selection. 
Second, the SINR constraints (\ref{P_SINR}) are non-convex, in which the beamforming vectors (i.e., $\{ \bm \omega_k \}$) and the phase-shift matrices (i.e., $\{ \bm \Theta_l \}$) are coupled. 
Third, the constant modulus constraints (\ref{P_unit}) are non-convex. 
To tackle the aforementioned challenges, we shall propose an efficient algorithm, which alternately solves SOCP and MIQP problems, as presented in the following section. 


\section{Alternating SOCP and MIQP Algorithm}
In this section, we propose an alternating optimization algorithm to solve the network power minimization problem. 
In each alternating procedure, we optimize the BS beamforming vectors by solving an SOCP problem and the active RIS set as well as the corresponding RIS phase-shift matrices by solving an MIQP problem. 
In particular, the MIQP problem is transformed into an SDP problem by applying the binary relaxation and SDR techniques. 


\subsection{SOCP for BS Beamforming Optimization}
With given active RIS set $\mathcal{A}$ and phase-shift matrices $\{ \bm \Theta_l \}$ of the active RISs, the composite channel responses of the direct and reflect links between the BS and each user are fixed. 
For notational ease, we define $\bm{\tilde{h}}_{k}^{\mathrm{H}} (\mathcal{A})\in \mathbb{C}^{1 \times M}$ as 
\begin{eqnarray} \label{Eq_CompH}
\bm{\tilde{h}}_{k}^{\mathrm{H}} (\mathcal{A}) = \sum_{l \in \mathcal{A}} \bm{h}_{l,k}^{\mathrm{H}} \bm{\Theta}_{l} \bm{G}_{l} + \bm g_k^{\mathrm{H}}. 
\end{eqnarray}

Based on (\ref{Eq_SINRk3}) and (\ref{Eq_CompH}), the SINR constraints in (\ref{P_SINR}) can be equivalently rewritten as
\begin{eqnarray}\label{Eq_SINRSOC01}
{\left|\bm{\tilde{h}}_{k}^{\mathrm{H}}(\mathcal{A}) \bm \omega_k \right|^2} \geq \gamma_k \left({\sum_{j \neq k} \left|\bm{\tilde{h}}_{k}^{\mathrm{H}}(\mathcal{A}) \bm \omega_j \right|^2 + \sigma_k^2} \right), \forall \, k \in \mathcal{K}.
\end{eqnarray}
%
By extracting the hidden convexity, constraints (\ref{Eq_SINRSOC01}) can be reformulated as the following second order cone (SOC) constraints \cite{6832894}
\begin{eqnarray}\label{Eq_SINRSOC}
\mathcal{C}_1(\mathcal{A}): \hspace{-6mm} && \frac{1}{\sqrt{\gamma_{k} \sigma_k^{2}}} \Re\left(\bm{\tilde{h}}_{k}^{\mathrm{H}}(\mathcal{A}) \bm \omega_k\right) \! \geq \! \sqrt{\sum_{j \neq k} \frac{1}{\sigma_k^{2}}\left|\bm{\tilde{h}}_{k}^{\mathrm{H}}(\mathcal{A}) \bm \omega_j\right|^{2} \!+ \! 1}, \nonumber \\
\hspace{-6mm} && \hspace{55mm}\forall \, k \in \mathcal{K}, 
\end{eqnarray}
where $\Re(\cdot)$ denotes the real part of a complex number. 

When the active RIS set and the corresponding phase-shift matrices are given, the network power minimization problem $\mathcal{P}$ can be reduced to the following optimization problem 
\begin{subequations} \label{OP_Omega}
\begin{align}
\mathrm{SOCP}: \quad \mathop{\text{minimize}}_{ \{ \bm \omega_k \}} \hspace{5mm}  & \sum_{k \in \mathcal{K}} \frac{1}{\eta} \| \bm \omega_k \|^2  \\
\text{subject to} 
\hspace{5mm}  & \sum_{k \in \mathcal{K}} \| \bm \omega_k \|^2 \le P_{\mathrm{max}}, \\ 
\hspace{5mm}  &  \mathcal{C}_1(\mathcal{A}). 
\end{align}
\end{subequations}
Note that problem (\ref{OP_Omega}) is an SOCP problem, which can be solved by the existing solvers such as CVX \cite{cvx}. 

\subsection{MIQP for Active RIS Set and Phase Shifts Optimization}
With given beamforming vectors $\{ \bm \omega_k \}$, problem $\mathcal{P}$ is reduced to the joint optimization of the active RIS set $\mathcal{A}$ and the corresponding phase-shift matrices $\{ \bm \Theta_l \}$, given by 
\begin{subequations}  \label{OP_Theta}
\begin{align}
\mathop{\text{minimize}}_{ \mathcal{A},\{ \bm \Theta_l \}} \hspace{3mm}  & \sum_{l \in \mathcal{A}} P_{\mathrm{RIS}}(N_l)  \\
\text{subject to} 
\hspace{3mm}  & \mathrm{SINR}_k (\mathcal{A}) \ge \gamma_k, \forall \, k \in \mathcal{K},  \\
 \hspace{3mm}  & | \theta_{l,n}| = \rho_l, \forall \, l \in \mathcal{A},\forall \, n \in \mathcal{N}_l. 
\end{align}
\end{subequations}

To address the combinatorial challenge, we introduce an auxiliary binary variable $a_l \in \{0,1\}$ to indicate whether or not RIS $l$ is active, $\forall \, l \in \mathcal{L}$. 
In particular, we denote $a_l = 1$ if RIS $l$ is active, and $a_l = 0$ otherwise. 
For notational ease, we further denote $\bm a = \{a_1, a_2, \ldots, a_L\} \in \{0,1\}^L$. 
To facilitate the problem transformation, we denote $v_{l,n} = a_l \theta_{l,n},\forall \, l \in \mathcal{L}, \forall \, n \in \mathcal{N}_l$, and $\bm{v}_l = [v_{l,1},v_{l,2},\ldots,v_{l,N_l}]^\mathrm H \in \mathbb{C}^{N_l \times 1}, \forall \, l \in \mathcal{L}$. 
With given beamforming vectors $\{ \bm \omega_k \}$, $ b_{k,j}=\bm g_{k}^{\mathrm{H}}\bm \omega_j $ and $\bm c_{k,j}(l)=\mathrm {diag}(\bm h_{l,k}^{\mathrm{H}})\bm G_l \bm \omega_j$ are fixed, $\forall \, k, j \in \mathcal{K}, \forall \, l \in \mathcal{L}$. 
Hence, we have $ \bm v_l^{\mathrm{H}}\bm c_{k,j}(l)= a_l \bm h_{l,k}^{\mathrm{H}} \bm \Theta_l \bm G_l \bm \omega_j, \forall \, k, j \in \mathcal{K}, \forall \, l \in \mathcal{L}$. 
With these notations, the SINR constraints (\ref{P_SINR}) can be rewritten as 
\begin{eqnarray} \label{Eq_SINRk}
\hspace{-16mm} && \gamma_k \left (\sum_{j \neq k} | \sum_{l \in \mathcal{L}} \bm v_l^{\mathrm{H}}\bm c_{k,j}(l) + b_{k,j} |^2 + \sigma_k^2 \right) \nonumber\\
\hspace{-16mm} && \hspace{3mm} \leq \left| \sum_{l \in \mathcal{L}} \bm v_l^{\mathrm{H}}\bm c_{k,k}(l) + b_{k,k} \right|^2, \forall \, k\in \mathcal{K}.
\end{eqnarray}

To address the non-convexity of constraint (\ref{Eq_SINRk}), we further denote $\bm{\tilde{c}}_{k,j}=[\bm c_{k,j}(1)^{\mathrm{H}},\bm c_{k,j}(2)^{\mathrm{H}},\dots,\bm c_{k,j}(L)^{\mathrm{H}}]^{\mathrm{H}} \in \mathbb{C}^{{\widehat{N}} \times 1},\forall k,j \in \mathcal{K}$, and $\bm{\tilde{v}}=[\bm{v}_1^{\mathrm{H}},\bm{v}_2^{\mathrm{H}},\dots,\bm{v}_L^{\mathrm{H}}]^{\mathrm{H}} \in \mathbb{C}^{{\widehat{N}} \times 1}$, where ${\widehat{N}=\sum_{l \in \mathcal{L}}N_l}$.
As a result, we have $\sum_{l \in \mathcal{L}}\bm v_l^{\mathrm{H}}\bm c_{k,j}(l)={\bm{\tilde{v}}}^{\mathrm{H}}{\bm{\tilde{c}}}_{k,j}, \forall \, k, j \in \mathcal{K}$. 
By introducing an auxiliary variable $t$, constraint (\ref{Eq_SINRk}) can be rewritten as 
\begin{eqnarray}
\mathcal{C}_2(\mathcal{L}): \hspace{-6mm} && \gamma_k \left(\sum_{j \neq k} \left( \bm{q}^{\mathrm{H}}\bm D_{k,j} \bm{q} + |b_{k,j}|^2\right)  + \sigma_k^2 \right) \nonumber \\
\hspace{-6mm} && \hspace{3mm} \leq  \bm{q}^{\mathrm{H}}\bm D_{k,k} \bm{q} + |b_{k,k}|^2, \forall \, k \in \mathcal{K},
\end{eqnarray}
where 
\begin{eqnarray}
\boldsymbol{D}_{k,j}=
\left[
\begin{array}{ll}{{\bm {\tilde{c}}}_{k,j} \bm {\tilde{c}}^{\mathrm{H}}_{k,j}} & {{\bm {\tilde{c}}_{k,j}} b_{k,j}^{\mathrm{H}}} \\ 
{b_{k,j}{\bm {\tilde{c}}}_{k,j}^{\mathrm{H}}}
& {0}\end{array}\right] \in \mathbb{C}^{(\widehat{N}+1) \times (\widehat{N}+1)}, 
\end{eqnarray}

\begin{eqnarray}
\bm{{q}}=\left[\begin{array}{l}\bm{\tilde{v}} \\ {t}
\end{array}
\right]\in \mathbb{C}^{(\widehat{N}+1) \times 1},
\end{eqnarray}
and $\mathcal{I}_l = \{ \sum_{j=1}^{l-1}N_j + 1, \dots, \sum_{j=1}^{l}N_j \}, \forall \, l \in \mathcal{L}$.

By introducing the auxiliary binary variable $\bm a$, we replace the active RIS set $\mathcal{A}$ with the complete RIS set $\mathcal{L}$, and problem (\ref{OP_Theta}) can be equivalently reformulated as follows
\begin{subequations} \label{OP_act}
	\begin{align}
\mathrm{MIQP}:	\mathop{\text{minimize}}_{ \bm a,  \bm q }  \hspace{3mm} &\sum_{l \in \mathcal{L}} a_l^2 P_{\mathrm{RIS}}(N_l)  \\
	\text{subject to} \hspace{6mm}
	\label{OP_act01}  \hspace{-3mm} & \mathcal{C}_2(\mathcal{L}),  \\
\label{OP_act_q1}\hspace{-3mm}  & |\bm q_i|^2 = a_l^2 \rho_l^2, \forall \, l \in \mathcal{L}, \forall \, i \in \mathcal{I}_l, \\
\label{OP_act_q2}\hspace{-3mm}  & |\bm q_{\widehat{N}+1}| = 1, \\
	\label{OP_act_a} \hspace{-3mm} &  a_l^2\in\{ 0,1\},\forall \, l \in \mathcal{L}. 
	\end{align}
\end{subequations}
%
%
%

We relax the non-convex binary constraints (\ref{OP_act_a}) as the unit interval constraints \cite{lee2011mixed}, thereby yielding the following homogeneous non-convex QCQP problem \cite{So2007On}
\begin{eqnarray} \label{OP_NC}
\mathop{\text{minimize}}_{ \bm a,  \bm q }  \hspace{-3mm} &&\sum_{l \in \mathcal{L}} a_l^2 P_{\mathrm{RIS}}(N_l)  \nonumber \\
	\text{subject to} \hspace{0mm}
  \hspace{-3mm} && \mathcal{C}_2(\mathcal{L}), \nonumber \\
	\hspace{-3mm}  && \mathrm{constraints}\; (\ref{OP_act_q1}), (\ref{OP_act_q2}),\nonumber \\
 \hspace{-3mm}  && 0\leq a_l^2 \leq 1, \forall \, l  \in \mathcal{L}. 
\end{eqnarray}

To tackle this issue, we adopt the powerful SDR technique, which reformulates problem (\ref{OP_NC}) into the rank constrained matrix optimization problem via matrix lifting, followed by dropping the rank-one constraint. We denote positive semidefinite (PSD) matrix $\bm Q= {\bm q} {\bm q}^{\mathrm{H}} \in \mathbb{C}^{(\widehat{N}+1) \times (\widehat{N}+1)}$ with $\bm Q \succeq 0$ and ${\rm {rank}}(\bm Q)=1$. 
By further denoting $ \widehat{a}_l = a_l^2, \forall \, l \in \mathcal{L} $, $ \widehat{\bm a} = \{ \widehat{a}_1,\widehat{a}_2,\dots,\widehat{a}_L \} $, $\mathrm{Tr}(\bm D_{k,j} \bm Q) = {\bm q}^{\mathrm{H}}\bm D_{k,j} {\bm q}$ and dropping the rank-one constraint, problem (\ref{OP_NC}) can be relaxed as the following SDP problem
%
%
\begin{eqnarray} \label{OP_The}
\mathop{\text{minimize}}_{ \widehat{\bm a}, \bm Q} & & \hspace{-3mm} \sum_{l  \in \mathcal{L}} \widehat{a}_l P_{\mathrm{RIS}}(N_l) \nonumber \\
\text{subject to} & & \hspace{-3mm}\gamma_k \sum_{j \neq k}{\rm Tr}(\bm D_{k,j}\bm Q) + \gamma_k \left( \sum_{j \neq k}|b_{k,j}|^2+ \sigma_k^2 \right)\nonumber\\
&& \hspace{-1mm} \leq  {\rm Tr}(\bm D_{k,k}\bm Q) + |b_{k,k}|^2, \forall \, k\in \mathcal{K},\nonumber \\
&& \hspace{-3mm} \bm Q(i,i) = \widehat{a}_l \rho_l^2, \forall\, l \in \mathcal{L}, \forall\, i \in \mathcal{I}_l,\nonumber\\
&&\hspace{-3mm} \bm Q(\widehat{N}+1,\widehat{N}+1) = 1, \nonumber\\
&& \hspace{-3mm} \bm Q \succeq 0,\nonumber\\
&& \hspace{-3mm} 0\leq \widehat{a}_l \leq 1, \forall \, l \in \mathcal{L}.
\end{eqnarray}

The resulting SDP problem can be efficiently solved by the existing solvers such as CVX \cite{cvx}.
We denote $\bm{Q}^{*}$ and $\bm{q}^{*}$ as the solution of problems (\ref{OP_The}) and (\ref{OP_NC}), respectively. If $\bm{Q}^{*}$ is rank-one, $\bm{q}^{*}$ can be recovered by using Cholesky decomposition on $\bm{Q}^{*}$. If $\bm{Q}^{*}$ fails to be rank-one, $\bm{q}^{*}$ can be obtained by using Gaussian randomization on $\bm{Q}^{*}$ \cite{5447068}. 
After obtaining the solution $\bm q^*$, we can recover $\tilde {\bm v}^*=[\bm q^*/\bm q^*_{\widehat{N}+1}]_{(1:\widehat{N})}$, where $[\bm f]_{(1:\widehat{N})}$ denotes the vector that contains the first $\widehat{N}$ elements in $\bm f$. 
By splitting $\tilde {\bm v}^*$, we obtain the set $\{ \bm v_1^*,\ldots, \bm v_L^*\}$, which can be used to recover $\bm \Theta_{l}^* = \mathrm{diag}((\bm v_{l}^*)^{\mathrm{H}}), \forall \, l \in \mathcal{L}$. Besides, the auxiliary binary variable set, denoted as $\widehat{\bm a}^*$, can be obtained to recover the active RIS set $\mathcal{A}^*=\{ l \,| \,\widehat{a}_{l}^*=1, \forall \, l \in \mathcal{L}\}$.
%
%
The proposed algorithm to solve problem $\mathcal{P}$ is summarized in Algorithm \ref{Alg}, which solves problems (\ref{OP_Omega}) and (\ref{OP_The}) alternately in an iterative manner until convergence.

\begin{algorithm}[t] 
	\caption{Proposed Algorithm for Solving Problem $\mathcal{P}$}\label{Alg}
	\KwIn{Initial $\{\bm \Theta_l^0 \}$ and threshold $\epsilon$.}
	\For{t=1,2,\dots}
	{
		Given $\{\bm \Theta_l^{t-1} \}$, solve 	problem (\ref{OP_Omega}) to obtain $\{\bm \omega_k ^t\}$. \\
		Given $\{\bm \omega_k^t \}$, solve problem (\ref{OP_The}) to obtain $\widehat{\bm a}^t$ and $\bm Q^t$, and denote the solution as $\bm q^t$ after using Gaussian randomization on $\bm Q^t$. \\
		
		Sort $\widehat{\bm a}^t$ in the ascending order as $\tilde{\bm a}^t$, where $\tilde{\bm a}^t_{\pi_ 1} \leq \tilde{\bm a}^t_{\pi_ 2}\leq \dots \leq \tilde{\bm a}^t_{\pi_ L}$.\\
		Initialize $J_{low}=0,J_{up}=L,J_0=0,flag=0$.\\
		\While{$J_{up}-J_{low}> 1$}
		{
			Set $J_0=\lfloor\frac{J_{up}+J_{low}}{2}\rfloor$.\\
			Set $\tilde{\bm a}^t_{j}=0,\forall j\in \{1,\dots,J_0\}$, $\tilde{\bm a}^t_{j}=1,\forall j\in \{J_0+1,\dots,L\}$.\\
			Set $|\bm q^t_{\widehat{N}+1}| \!=\! 1, |\bm q^t_i| \!=\! \tilde{\bm a}^t_{\pi_ l}  \rho_{\pi_ l},\! \forall \, \pi_ l \!\in\! \mathcal{L}, \!\forall \, i \!\in\! \mathcal{I}_{\pi_ l}$.\\
			Check the feasibility of problem (\ref{OP_act}).\\
			\eIf{problem (\ref{OP_act}) is feasible}
			{
				Recover $\tilde {\bm v}^t\!=\![\bm q^t\!/\!\bm q^t_{\widehat{N}+1}]_{(1:\widehat{N})}$, split $\tilde {\bm v}^t$ to get $\{ \bm v^t_1,\!\dots\!, \bm v^t_L\}$,
				$\mathcal{A}^t\!=\!\{ \pi_ l | \tilde{\bm a}^t_{\pi_ l}\!=\!1, \!\forall\! \, \pi_ l \!\in\! \mathcal{L}\}$, and $\bm \Theta^t_{l} = \mathrm{diag}\left( (\bm v^t_{ l})^{\mathrm{H}}\right), \forall \, l \in \mathcal{A}^t$.\\
				Set $J_{low}=J_0,flag=1$.
			}
		    {Set $J_{up}=J_0$.} 
		}
		\If { The decrease of the network power consumption is below $\epsilon$ or flag=0}
		{ $\textbf {break}$. }	
	}
	\KwOut{ $\mathcal{A}^t, \{ \bm \omega_k ^t\}, \{ \bm \Theta_l ^t\}$ }
	
\end{algorithm}

\section{Numerical Results}
In this section, we present the numerical results of the proposed algorithm in a multi-RIS-assisted MISO cellular network. 
We consider a three-dimensional (3D) coordinate system consisting of one BS and $L=3$ RISs. 
The BS, located at $(0,0,50)$ meters, is equipped with $M = 10$ antennas. 
The RISs are deployed at $(0,40,40)$, $(40,60,40)$, and $(60,20,40)$ meters. 
Each RIS has $N_l = 12$ reflecting elements, $\forall \, l \in \mathcal{L}$. 
In addition, $K=6$ users are uniformly distributed in the range of $(0,0,0)\times (10,10,0)$ meters. 
All the channels suffer from both path loss and Rayleigh fading. 
As in \cite{wu2018intelligent, huang2019reconfigurable}, we set the path loss exponents of the BS-RIS link, the RIS-user link, and the BS-user link as 2.5, 2.4, and 3.5, respectively. 
Without loss of generality, we assume that all the users have the same SINR threshold, i.e., $\gamma_k=\gamma, \forall \, k \in \mathcal{K}$. Unless specified otherwise, we set the target data rate $R_k^{\min} = {\rm {log}(1+\gamma_k)}=2$ bits per channel use, $\sigma_k^2=-40$ dBm, $\forall \, k \in \mathcal{K}$, the drain efficiency $\eta=0.6$, the amplitude reflection coefficient $\rho _l=1,\forall \, l \in \mathcal{L}$, the maximum transmit power $P_{\max}=1000$ mW, and the circuit power of each RIS element $P_{\mathrm{RE}}=10$ mW \cite{huang2019reconfigurable}. 

To illustrate the effectiveness of the proposed algorithm, we consider two baseline schemes, termed as all-RIS-active and exhaustive search, respectively. 
For the all-RIS-active method, all the RISs are active and only the BS transmit power is minimized. 
For the exhaustive search method, all possible combinations of active RISs are checked and the algorithm complexity grows exponentially with the number of RISs. 

%

\begin{figure}[t] \centering
	\includegraphics[scale = 0.5]{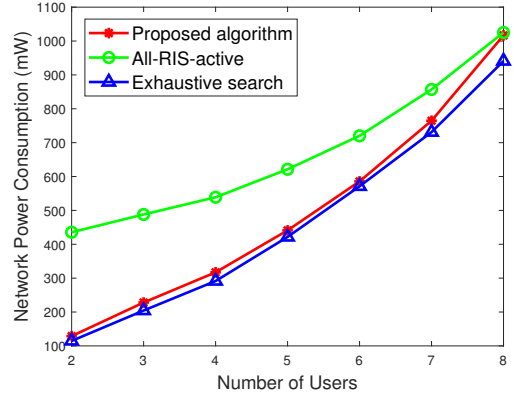}
	\caption{Network power consumption versus the number of users.}
	\label{Fig_K}
	\vspace{-5mm}
\end{figure}

Fig. \ref{Fig_K} shows the impact of the number of users (i.e., $K$) on the network power consumption of all algorithms under consideration. 
By dynamically switching off some RISs, the network power consumption of the proposed algorithm is much lower than that of the all-RIS-active method. 
In addition, the performance gap between the proposed algorithm and the exhaustive search algorithm is small, which demonstrates the effectiveness of the proposed algorithm. 
As the number of users increases, the traffic load in the network also increases. As a result, more network power consumption due to a larger number of active RISs and/or a higher BS transmit power is required to support the QoS requirements of all users.

\begin{figure}[t] \centering
	\includegraphics[scale = 0.49]{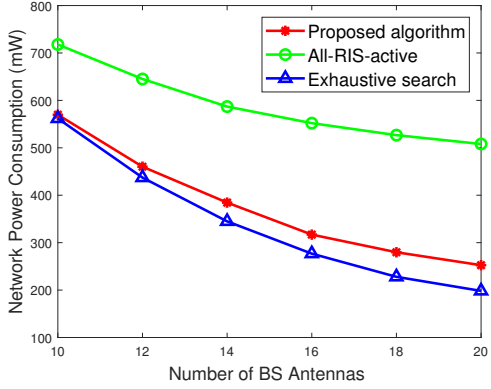}
	\caption{Network power consumption versus number of BS antennas. }
	\label{Fig_M}
	\vspace{-5mm}
\end{figure}

Fig. \ref{Fig_M} illustrates the impact of the number of BS antennas (i.e., $M$) on the network power consumption. 
As the value of $M$ increases, the BS transmit power of the all-RIS-active method decreases due to the increased power gain. 
The reduction of the BS transmit power can also be achieved by reducing the number of active RISs. 
By jointly optimizing the reduction of both the BS transmit power and the RIS circuit power, the decreasing rate of the proposed algorithm is larger than that of the all-RIS-active method. 

\begin{figure}[t] \centering
	\includegraphics[scale = 0.49]{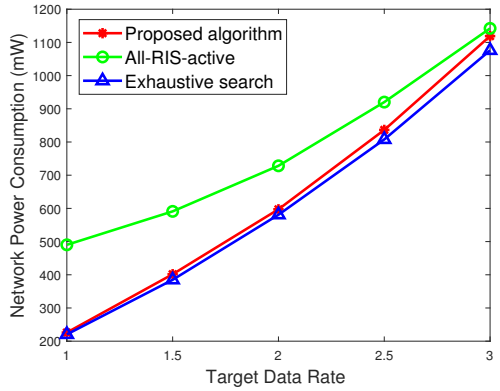}
	\caption{Network power consumption versus target data rate.}
	\label{Fig_Rate}
	\vspace{-5mm}
\end{figure}

Fig. \ref{Fig_Rate} shows the impact of the target data rate (i.e., $R_k^{\min},\forall \, k \in \mathcal{K}$) on the network power consumption. 
When the target data rate is low, the proposed algorithm has a significant performance gain over the all-RIS-active method, and achieves almost the same performance as the exhaustive search method. 
As the target data rate increases, more RISs are required to be active to meet the target data rate requirements, and hence the performance gap between the proposed algorithm and the all-RIS-active method reduces.


%

\section{Conclusions}
\label{Sec_Con}
In this paper, we formulated a new RIS selection and beamforming optimization problem to minimize the network power consumption of a multi-RIS-assisted multiuser MISO system. 
To account for the tradeoff between the BS transmit power consumption and the RIS circuit power consumption, we proposed the joint design of the beamforming vectors at the BS, the active RIS set, and the phase-shift matrices at the active RISs. 
To solve the challenging MIQP problem, we developed an efficient algorithm that alternately solves the SOCP and MIQP problems, where the MIQP problem was transformed to an SDP problem by applying binary relaxation and SDR. 
Simulation results showed the effectiveness of the proposed algorithm in reducing the network power consumption and demonstrated that the RIS circuit power cannot be ignored when designing green RIS-assisted wireless networks. 


\ifCLASSOPTIONcaptionsoff
  \newpage
\fi

\bibliographystyle{IEEEtran}
\bibliography{ref}

\end{document}